\documentclass[prb,twocolumn,preprintnumbers,amsmath,amssymb]{revtex4}

\usepackage{graphicx}
\usepackage{dcolumn}
\usepackage{bm}
\usepackage{amsmath}
\usepackage{amssymb}
\usepackage{amsthm}
\usepackage{amsfonts}
\usepackage{enumerate}
\usepackage{latexsym}

\begin{document}

\title {Rarita-Schwinger-Weyl semimetal in $J_{eff}=3/2$ electron systems}

\author{ Long Liang$^{1,2}$ and Yue Yu$^{1,2}$}

\affiliation {1. Department of Physics, Center for Field
Theory and Particle Physics, State Key Laboratory of Surface Physics and Collaborative Innovation Center of
 Advanced Microstructures, Fudan University, Shanghai 200433,
China \\2. State Key Laboratory of Theoretical Physics, Institute of
Theoretical Physics, Chinese Academy of Sciences, P.O. Box 2735,
Beijing 100190, China 
}

\begin{abstract}
We propose a relativistic $J_{eff}=3/2$ semimetal with 4$d^1$ or 5$d^1$ electrons on a cubic lattice when the strong spin-orbital coupling takes over the Hunds' coupling. A relativistic spinor with spin 3/2  is historically called Rarita-Schwinger spinor. In the massless case,  the  right- and left-handed chiral degrees of freedom of the Rarita-Schwinger spinors are independent. In the lattice model that we propose, the right- and left- handed gapless points in Brillouin zone are separated. We call this linearly dispersed semimetal Rarita-Schwinger-Weyl semimetal, similar to Weyl semimetal for spin 1/2 systems.  There is a network of gapless Fermi arcs in the surface Brillouin zone if $n_1+n_2+n_3$ is even for the normal vector $(n_1,n_2,n_3)$ of the surface while the surface is insulator if $n_1+n_2+n_3$ is odd.

\end{abstract}


\maketitle

\noindent{\it Introductions.} In 1936, eight years after derived his famous relativistic electron equations of motion, Dirac generalized these equations to higher spin relativistic particles \cite{dirac}.  The first important example was the  recovering of Maxwell equations.
The next simplest particle except the electron and the photon is of spin-3/2 and obeys so-called Rarita-Schwinger (RS)  equations\cite{RS}. This fermonic field later played an important role in supergravity theory, known as the super-partner of the graviton, the gravitino \cite{sg}.  

Electrons and photons are particles accompanying us on daily life while the  RS particles are not found even in experiments of  high energy physics or in cosmology observations.  Recently, the interplaying between high energy physics and condensed matter physics supplies a new playground to the relativistic systems, e.g., Dirac semimetal in graphene \cite{graphene} and three dimensions \cite{dirac1,dirac2,dirac3},  topological insulator \cite{ti,ti1}, supersymmetric systems \cite{lee,yy,gdv}, and newly proposed \cite{weyl} and discovered \cite{weyl1,weyl2,weyl3,weyl4,weyl5,weyl6,weyl7,weyl8}  Weyl semimetal. Can we expect  a RS semimetal in this playground?

The earlier theoretical prediction for Weyl semimetal was based on the $J_{eff}=1/2$ states in iridates \cite{weyl} which belongs to a large class of 4$d$ and 5$d$ transition metal oxides with a strong spin-orbital coupling.  When the $d$-orbitals are below 2/3 filling, the crystal field projects the electron configurations to the $t_{2g}$ orbitals. When the strong spin-orbital coupling dominates, the electrons fill the $J_{eff}=1/2$ or $J_{eff}=3/2$ depending on the filling factor. For iridates, the 5$d^5$ electrons occupy all $J_{eff}=3/2$ states and half fill $J_{eff}=1/2$ orbitals \cite{rev}.  For $d^{1,2}$ electrons, they quarterly or half occupy the $J_{eff}=3/2$ states. The representatives of these $J_{eff}=3/2$ materials are the ordered double perovskites with the chemical formula $A_2B'BO_6$ where $B'$ ions are commonly 4$d^{1,2}$ and 5$d^{1,2}$ transition metals', e.g. Mo$^{+5}$,Re$^{+6}$,Os$^{+7} $ for $d^1$ and  Re$^{+5}$,Os$^{+6}$ for $d^2$. Several exotic magnetic phases and a quadrupolar phase were presented in these strongly correlated materials\cite{rev,gchen1,gchen2}.


Due to the particle-hole symmetry, the relativistic physics usually emerges in a half-filled lattice model. For the massive RS theory, this implies a 4$d^2$ or 5$d^2$ system is required. However, the above strongly correlated systems are not likely because the first Hund's rule due to the interaction makes two electrons first form a total spin ${\bf S}=1$ state which then couples to ${\bf l}=1$ state. The final effective state is projected to $J_{eff}=2$ \cite{gchen2}. 
On the field theoretical side, there are also many complexes for the massive RS field itself. We thus restrict our attention to  the massless RS field. 

The massless relativistic spinor systems with any spin possess only two physical degrees of freedom, the highest and lowest  helical states \cite{lurie}. Maxwell field with only transverse components is the well-known example.  The solutions of the massless RS equations possess the helicity $\pm3/2$ associated with the right- and left-handed chirality. The helicity $\pm1/2$ states are not zero mass states. 
Therefore, the fourfold-degeneracy of the $J_{eff}=3/2$  bands in the transition metal compound where the massless RS field emerges 
 needs to be lift: Two helicity $\pm1/2$ bands are gapped from the helicity $\pm3/2$ bands.  The relativistic theory emerges from the half-filling of  two lower bands, i.e., the quarter filling  $J_{eff}=3/2$.  We then see that appropriate systems are  4$d^1$ or 5$d^1$ transition metal compounds. They are semi metallic, similar to the Dirac and  Weyl semimetals.  We call these semi-metallic states  {\it the RS or RS-Weyl semimetal}. We will present a simplest lattice model where the RS field emerges while searching for realistic materials is not our goal at this stage. 

On the other hand,  we will not concern the Lagrangian presented in Rarita and Schwinger's original paper in 1941 \cite{RS}.  A modern version for the  gravitino will be used because of its simplicity and gauge invariance \cite{sg}.  The latter is also that high energy theorists are studying and well-behaved in an external field \cite{adler}. 

To our knowledge, the RS physics was not concerned in the condensed matter physics context. We therefore will  briefly introduce the RS theory and make the convention to be familiar with the readers in condensed matter physics area. We then construct a tight-binding  model for $d_{xy},d_{yz}$ and $d_{zx}$ electrons on a cubic lattice where $d_{xy}$ refers to the $d$-electron annihilation operator with the $xy$-orbital component, etc.  The long wave length effective theory shows that this is a RS-Weyl semimetal with eight chiral Weyl points. The surface states are dependent on what surface of the system is. If the normal vector $(n_1,n_2,n_3)$ of the surface Brillouin zone is even, i.e., $ n_1+n_2+n_3$ is even, there is a network of the gapless Fermi arcs. For a surface Brillouin zone with an odd normal vector, the surface is insulating. The example of the former is the surface $[110]$ while $[001]$ and $[111]$ are the representatives of the latter.


\vspace{2mm} 

\noindent{\it Rarita-Schwinger theory.} The RS equations for a sixteen component vector-spinor field $\psi_{\mu\alpha}$ in 3+1 dimensions are given by
\begin{eqnarray}    
&&(i\gamma^\mu\partial_\mu-m)\psi_\nu=0,\label{rs1}\\
&&\chi=\gamma^\mu\psi_\mu=0,\label{rs2}
\end{eqnarray} 
where the convention we use are: $\mu=0,1,2,3$  denote the time-space indices with flat metric $\eta^{00}=\eta_{00}=1$ and $\eta_{ii}=-\eta^{ii}=-1$; $\alpha=s\sigma$ for $s=R,L$ and $\sigma=\uparrow,\downarrow$ are chiral and spin indices, respectively.  The Weyl's  gamma matrices are chosen:
  \begin{eqnarray}
&&\gamma^0=\left(\begin{array}{cccccccc}
0&I\\
I&0\end{array}\right),
\gamma^i=\left(\begin{array}{cccccccc}
0&\sigma^i\\
-\sigma^i&0\end{array}\right),
\gamma^5=\left(\begin{array}{cccccccc}
-I&0\\
0&I\end{array}\right).
\nonumber
\end{eqnarray}
In Eq. (\ref{rs2}), the four vector indices are  contracted over so that $\chi$ is a pure Dirac spinor. $\chi=0$ projects out the spin-1/2 sector and leaves only the degrees of freedom of the spin-3/2 sector.   A non-zero mass $m$ brings many problems to the RS system. It was known that if $m\ne 0$, there will be  fermonic modes with superluminal velocities if the RS field couples to the external electromagnetic field in a minimal way \cite{mass1,mass2}. However, this does not cause a trouble for the condensed matter system because it is allowed that the velocity of a collective mode exceeds the 'speed of light", the Fermi velocity.

 The bad thing for us is that
the four vector leads to a superfluous component $\psi_0$ if we identify $i=1,2,3$ to the orbital degrees of freedom $xy,yz,zx$ of the $d$-electron. One may use Eq. (\ref{rs2}) to eliminates $\psi_0$. The canonical quantization procedure is very inconvenient  because the Hamiltonian after eliminating $\psi_0$ becomes very complicated \cite{hamil}, although a quantized free RS field is well defined \cite{lurie}. Thus, it is difficult to realize a massive RS field with  a simple model. 
We will study a massless RS field in this work.   We take a simple form which used in supergravity for the gravitino RS  field. In a flat space-time, the RS Lagrangian reads \cite{sg}
\begin{eqnarray}
{\cal L}=-i\bar\psi_\mu\gamma^{\mu\nu\lambda}\partial_\nu\psi_\lambda\label{lag}
\end{eqnarray} 
where $\bar\psi_\mu=\psi^\dag_\mu\gamma^0$ and $\gamma^{\nu\mu\lambda}=\gamma^{[\mu}\gamma^\nu\gamma^{\lambda]}$ is the total antisymmetricized product of three gamma matrices.  It is obvious that the Lagrangian is gauge invariant under $\psi_\mu\to \psi_\mu+\partial_\mu\epsilon$ for an arbitrary spinor $\epsilon$. The equations of motion can be written in many equivalent ways \cite{equiv}, e.g, 
\begin{eqnarray}
\gamma^\mu\partial_\mu\psi_\nu-\partial_\nu\chi=0.\label{RS2}
\end{eqnarray}
in terms of Lagrangian (\ref{lag}).
Imposing the gauge fixing $\chi=0$, Eq.(\ref{RS2}) recovers the massless RS equations. It is also easy to have a supplementary condition $\partial_\mu\psi^\mu=0$ according to the RS equations. 

\vspace{2mm}

\noindent{\it $\psi_0=0$ gauge and two component chiral spinors.} Since $\psi_\mu$ is four component vector, it is still difficult to connect with a condescend system. The gauge invariance of Lagrangian (\ref{lag}) enables us to take $\psi_0=0$ gauge as taking $A_0=0$ gauge for the electromagnetic field. In this gauge,  the supplementary condition are $\partial_i\psi^i=0$, which also projects the system to ${\bf S}=3/2$. 

To more explicitly link to a condensed matter model, we use two component forms of the chiral RS spinor 
, i.e., $\psi_{i\alpha}=(c_{iL\sigma},c_{iR\sigma})^T$. Making Fourier transformations $
 c_{iR(L)\sigma}=\sum_pc_{R(L)\sigma p}e^{ip_\mu x^\mu}
 $ with $p_\mu=(\omega,{\bf p})$, the Lagrangian for $\psi_0=0$ is given by
\begin{eqnarray}
{\cal L}_{\omega,\bf p}&=& -\omega\sum_{s=L,R}c^\dag_{isp} c^i_{sp}
\nonumber\\
&+&\epsilon^{ijk}c^\dag_{iLp}p_jc_{kLp}-\epsilon^{ijk}c^\dag_{iRp}p_jc_{kRp}.\label{lop}
\end{eqnarray}
This Lagrangian is separated into the right- and left-handed chiral sectors as that for the massless spin-1/2 Dirac spinor.  
The supplementary conditions are $p^ic_{isp}=0$.   
 is a residual gauge invariance as $c_{is}(x)\to c_{is}(x)+\partial_i\epsilon({\bf r})$ if the spinor $\epsilon({\bf r})$ is independent of time. To fix the gauge invariance, we impose $\sigma^ic_{isp}=0$. 
The Hamiltonian then can be read out
\begin{eqnarray}
{\cal H}_{\bf p}&=&{\cal P}_{3/2}{\cal H}^{un}_{\bf p}{\cal P}_{3/2},\label{hp}\\
{\cal H}^{un}_{\bf p}&=&
-i\epsilon^{ijk}c^\dag_{iLp}p_jc_{kLp}+i\epsilon^{ijk}c^\dag_{iRp}p_jc_{kRp},
\end{eqnarray}
where ${\cal P}_{3/2}$ denotes the projection to the spin-3/2 sector.
The matrices of the left- and right handed unprojected Hamiltonians ${\cal H}^{un}_s$ are of a simple form
\begin{eqnarray}
{\cal H}^{un}_{R,L;6\times6}=\pm {\bf p}\cdot{\bf T}\otimes I_2 \label{hmatrix}
\end{eqnarray} 
where $(T^i)^{jk}=i\epsilon^{ijk}$ are three generators of SO(3) group and $I_2$ is $2\times 2$ unitary matrix for spin index.
 Diagonalizing $\mathcal{H}$ can be done by diagonalizing ${\cal H}^{un}_s$ and making an projection ${\cal P}_{3/2}$.   It is easy to see the eigen values of ${\cal H}^{un}_{L,R}$ are $\omega=0$ and $\omega=\pm|{\bf p}|$. The supplementary condition suppresses the eigen state corresponding to $\omega=0$.  The solutions of $\omega=\pm |{\bf p}|$ are not independent: The positive (negative) solution for the right-(left-)handed RS field is particle while the other is hole,or antiparticle. It is easy to check the right-(left-)handed solution is of the helicity 3/2 (-3/2) \cite{lurie,free}.  
 Therefore, projecting to $J_{eff}=3/2$, the genuine physical degrees of freedom for the ground states are 1+1. The electron operators can be expanded as
 \begin{eqnarray}
 c_{iR(L)\sigma}=\sum_p (c_{\pm3/2,p}{\cal U}_{R(L)i\sigma} e^{ipx}+d^\dag_{\pm3/2,p}{\cal V}_{R(L)i\sigma}e^{-ipx})\nonumber
  \end{eqnarray}
  with $\sigma^i{\cal U}_{R(L)i}=\sigma^i{\cal V}_{R(L)i}=0$.  The solution vector-spinors ${\cal U}$ and ${\cal V}$ are normalized by
  ${\cal U}^{i\dag}_L\sigma^j{\cal U}_{iR}={\cal V}^{i\dag}_L\sigma^j{\cal V}_{iR}=-p^j/\omega$ and ${\cal U}^{i\dag}_{R(L)}{\cal V}_{iR(L)}=0$.  ${\cal U}$ and ${\cal V}$ are not independent and they are related by the charge conjugation. 
  
 With the RS field operators $\tilde c_{\pm3/2,p}$ and $\tilde d^\dag_{\pm3/2,p}$  in the diagonalized basis, 
 \begin{eqnarray}
 c_{iR(L)\sigma}=\sum_p (\tilde c_{\pm 3/2,p}\tilde{\cal U}_{R(L)i\sigma} e^{ipx}+\tilde d^\dag_{\pm3/2, p}\tilde{\cal V}_{R(L)i\sigma}^he^{-ipx})\nonumber
  \end{eqnarray}
   with $\tilde\sigma^i\tilde{\cal U}_{R(L)i}=\tilde\sigma^i\tilde{\cal V}_{R(L)i}=0$ .  The diagonalized Hamiltonian then reads
  \begin{eqnarray}
  H=\sum_{{\bf p},\lambda=\pm3/2}|{\bf p}|(\tilde c^\dag_{\lambda p}\tilde c_{\lambda p}+\tilde d^\dag_{\lambda p}\tilde d_{\lambda p}).
  \end{eqnarray}

Defining a unit vector ${\bf n}={\bf p}/|{\bf p}|$ according to (\ref{hmatrix})  and after the projection, a topological winding number is explored for a chiral RS field
\begin{eqnarray}
\nu_{R,L}=\pm \int _{S^2} d^2p {\bf n}\cdot\frac{\partial {\bf n}}{\partial p_x}\times \frac{\partial {\bf n}}{\partial p_y} \label{wn}\end{eqnarray}
where $S^2$ is a two-sphere which wraps the Weyl point in the wave vector space. 

\vspace{2mm}

\begin{figure}[htb]
\vspace{-0.3cm}
\begin{center}
\includegraphics[width=10.0cm]{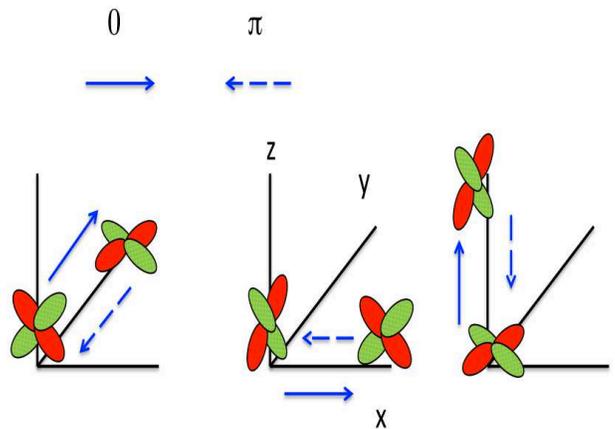}
\end{center}
\vspace{-1cm}
 \caption{\label{fig1} (color online) The hopping of electrons with different orbital components. There is a $\pi$ phase when electron hops in the negative direction.}
\end{figure}

\noindent{\it  Lattice model and RS-Weyl semimetal .} We now turn to the realization of RS field in condensed matter system.  As we mentioned,  we assume the 4$d^1$ or 5$d^1$ ions are on a simple cubic lattice for transition metal (e.g., Mo$^{+5}$,Re$^{+6}$, Os$^{+7}$ ) compounds. The crystal field projects the effective orbital angular momentum to ${\bf l}=1$. We consider the strong spin-orbital coupling takes over the Hunds' coupling, the $d^1$ electrons fill one of four $J_{eff}=3/2$ states, i.e., the quarter-filling of $J_{eff}=3/2$.  The Hamiltonian we are studying is as follows    
\begin{eqnarray}
H&=&-t\sum_{a} (\pm d_{xy;a}^\dag d_{zx;a\pm\delta_y}\pm d^\dag_{zx;a}d_{yz;a\pm\delta_x}\nonumber\\
&\pm& d^\dag_{yz,a}d_{xy,a\pm\delta_z})+h.c.-\lambda\sum_a {\bf l}_a\cdot {\bf S}_a+\cdots\label{lh}
\end{eqnarray}
where $\delta_{x,y,z}$ are the lattice vectors in the positive directions. The hopping term is between the different orbitals in different direction. Furthermore, the hopping in the negative $\delta_i$ direction carries a phase $\pi$ while carrying no phase in the positive direction (See Fig.\ref{fig1}). The third term is the spin-orbital coupling projected to $J_{eff}=3/2$ \cite{note1}. 
 The $\cdots$ refers to other hopping terms like  $t_d\sum_{i=xy,yz,zx}\sum_{\langle ab\rangle} d^\dag_{ia}d_{ib}$. We assume $t_d$ is small and neglect this term for the moment.

It is easy to check that the Hamiltonian is not invariant under space inversion $P$:  Under $d_{ia}\to -d_{i,-a}$,  $t$ term changes sign.  However, it is time reversal $T$ invariant: Under $d_{iR(L)\sigma,a}\to \sigma d_{iR(L),-\sigma,a}$, $H\to H$. Such properties under $P$ and $T$ resemble the recent observed Weyl semimetal in the  non-centrosymmetric transition metal monophosphides \cite{weyl1,weyl2,weyl3,weyl4,weyl5,weyl6,weyl7,weyl8}. The  role of the strong spin-orbital term is separating the $J_{eff}=1/2$ orbitals from $J_{eff}=3/2$ ones with an order of 1eV gap. Due to the minus sign before the spin-orbital coupling, the $J_{eff}=3/2$ states have a lower energy.  The effective Hamiltonian in the wave vector space is given by
\begin{eqnarray}
H&=&{\cal P}_{3/2}H^{un} {\cal P}_{3/2}, \label{wvh}\\
H^{un}&=&-2it\sum_{\bf k}\epsilon^{ijk}\sin k_j d^\dag_{i\bf k}d_{k\bf k}\nonumber
\end{eqnarray}
where the lattice spacing is taken to be one. 
Therefore, ${\bf k}=0$ and ${\bf k}=\vec\pi$ are  Weyl points for the left- and right handed chiral RS spinors by denoting $d_{i,0+\delta \bf k}=c_{iL,p}$ and $d_{i,\vec\pi+\delta \bf k}=c_{iRp}$. Hamiltonian (\ref{wvh}) exactly recovers the RS Hamiltonian (\ref{hp}) with $v_F=2t=1$. The helicity $\pm1/2$ states are projected over from the massless sector. 
Thus, Hamiltonian (\ref{lh}) describes {\it Rarita-Schwinger-Weyl semimetal}. 

\begin{figure}
\begin{center}
\includegraphics[width=7.0cm]{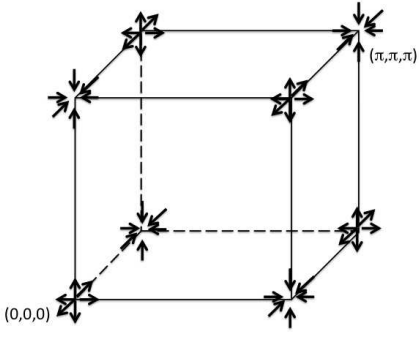}
\end{center}
 \caption{\label{fig2}  The singularity (monopole) configurations and distributions, as well as  the flows of the flux in Brillouin zone. The eight monopoles are located at $(0,0,0),  
 (\pi,0,0)$, $(0,\pi,0)$, $(0,0,\pi)$, $(\pi,\pi,0)$, $(\pi,0,\pi)$, $(0,\pi,\pi)$ and $(\pi,\pi,\pi)$ with the winding numbers $(1,-1,-1,-1,1,1,1,-1)$. }
\end{figure}

In fact, there are other three pairs of Weyl points which are located $\{(0,\pi,\pi),(\pi,0,0)\}, \{(\pi,0,\pi),(0,\pi,0)\}$ and $\{(\pi,0,0),(0,0,\pi)\}$ beside of the pair
$\{{\bf 0},\vec\pi\}$. In these points, there are singularities of the  vector field 
\begin{eqnarray}
{\bf n}=\frac{\sin {\bf k}}{\sqrt{\sum_j\sin^2 k_j}}.
\end{eqnarray}
Corresponding to these singularities, the winding numbers defined as Eq.(\ref{wn}) are $(1,-1)$ for each pair, respectively (See Fig. \ref{fig2}). These Weyl points can be thought of as the "monopoles" in Brillouin zone whose configuration is scratched in Fig. \ref{fig2}.

\begin{figure}
\begin{center}
\includegraphics[width=7.5cm]{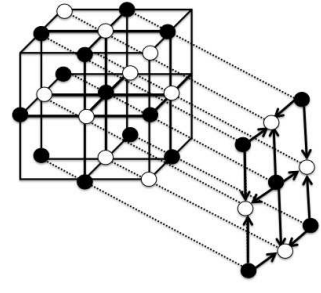}
\includegraphics[width=6.0cm]{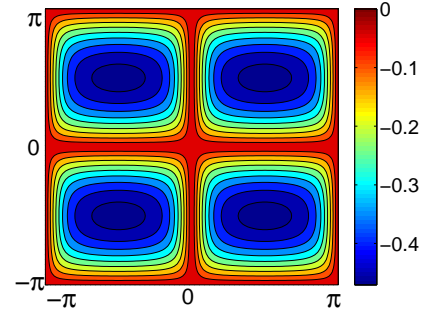}
\end{center}
 \caption{\label{fig3} (color online) The Fermi arc network of the surface states in the [110]-surface Brillouin zone. Upper: Images of the Weyl points with charges $\pm 1$ (black and white). Lower: The dispersions and the network of gapless Fermi arcs for $t=1$. }
\end{figure}

\noindent{\it Surface states and Fermi arc network.} One of the  characteristics of the Weyl semimetal for spin-1/2 system is the existence of Fermi ares in surface states\cite{weyl}, which also has been observed experimentally. We now study the surface state of the RS-Weyl semimetal. We consider a surface Brillouin zone with the normal vector along  the $(n_1,n_2,n_3)$ direction. The bulk Weyl points are projected onto this surface. A set of the Weyl points may have the same image. The charge of the image is defined by the sum of the winding numbers of these Weyl points in this set divided by the number of the points.  It is easy to see the charges of all images are zero if $n_1+n_2+n_3$ is odd while the charges of the images are $\pm 1$ if $n_1+n_2+n_3$ is even.  For a given surface, these images form a lattice. For a lattice where all sites are zero charged, there is no reason to see the Fermi arc connecting any pair of images while the Fermi arcs may appear between the positive and negative charged images in a charged site lattice. In Fig. \ref{fig3}(a)), we sketch the image network for n even surface, the [110] surface. The black and while spots correspond to the images with charge $\pm 1$.  We numerically calculate the surface states of the [001], [111] and [110] surfaces. We see that there is no gapless state for [001] and [111] while the dispersions of the surface states 
for the [110] are shown in Fig. \ref{fig3}(b).  For the latter,  as expected, the gapless Fermi arcs form a rectangular network and the images are the lattice sites of the network.        

\begin{figure}
\begin{center}
\includegraphics[width=6.0cm]{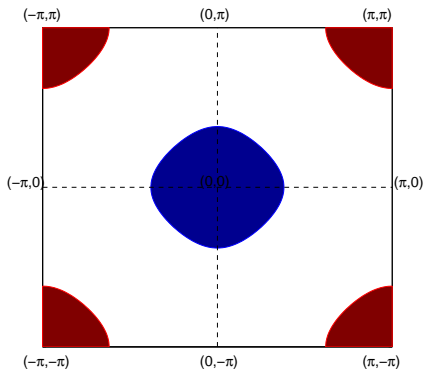}
\end{center}
 \caption{\label{fig4} (color online) The Fermi pockets of the surface states in the [001]-surface Brillouin zone for $t_d>t_\perp$. The blue pocket is particle's and the red one is hole's.}
\end{figure}

We now analytically verify the insulating nature of the odd surface. For example, we can show the insulating state in the [001] surface which  consists of two most-up coupled [001] layers, $(x,y, L/2)$ and $(x,y,L/2-\delta_z)$. The effective surface Hamiltonian is given by
 \begin{eqnarray}
 H_{001}&=&{\cal P}_{3/2} H^{un}_{s}{\cal P}_{3/2}\\
H^{un}_{001}&=&-t\sum_{a_x,a_y;\alpha=u,d} (\pm d_{xy;a,\alpha}^\dag d_{zx;a\pm\delta_y,\alpha}\nonumber\\
&\pm& d^\dag_{yz;a,\alpha}d_{zx;a\pm\delta_x,\alpha})\nonumber\\
&-&t_\perp\sum_{a_x,a_y}(-d^\dag_{yz,a,u}d_{xy,a,d}\\
&+&d^\dag_{yz,a,d}d_{xy,a,u})+h.c.\nonumber\\
&-&t_d\sum_{i,\alpha,\langle ab\rangle}d^\dag_{ia\alpha}d_{ib\alpha},
\nonumber \end{eqnarray}
 where $t_\perp$ is coupling between two layers and determined by the boundary condition; $\alpha=u,d$ are layer indices.
 We include the diagonal hopping term with the hopping amplitude $t_d$ which may affect the surface state.  
 One eigen value of $H^{un}_{001}$  is 
 $
 \omega=-t_d(\cos k_x+\cos k_y)
 $
 which is projected out after ${\cal P}_{3/2}$ projection. The real spectra of $H_{001}$ are given by
\begin{eqnarray}
\omega&=&-t_d(\cos k_x+\cos k_y)\\
&\pm& 2\sqrt{t_\perp^2+t^2\sin^2k_x+t^2\sin^2k_y}.\nonumber
\end{eqnarray}
 We assume $t_d<t$. When $t_\perp>t_d$, the surface state is insulating which is indeed that as we expected.  When $t_\perp=t_d$, the surface becomes a semimetal with quadratic dispersion at $(0,0)$ and $(\pi,\pi)$. If $t_\perp<t_d$, two Fermi pockets appear at $(0,0)$ and $(\pi,\pi)$. Surrounding $(0,0)$, it is a left-handed helical particle Fermi pokiest while another is hole's with a right helicity surrounding $(\pi,\pi)$ (See Fig. \ref{fig4}.)
 
We can also show that there is the Fermi arc network for an even surface by a similar analysis.  

 \noindent{\it Conclusions and perspectives. }  The spin
 -3/2 relativistic spinor state, the RS semimetal, was proposed in 
 transition metal compounds with $J_{eff}=3/2$ 4$d^1$ and 5$d^1$ electron configurations. The fruitful surface state properties were discussed and the network of the gapless Fermi arcs was predicted.  Our results point a new direction for the researches of topological states of matter after the topological states of spin-1/2  Dirac, Weyl and Majorana fermions are exhaustively studied.

 Unlike the spin-1/2 spinor whose properties are very well understood in field theory,  many properties of spin-3/2 relativistic field are not transparent. We were not aware of the topological classification of spin-3/2 spinor although the index theorem for the RS operator was known.    
The massive RS theory may also be corresponding to the topological insulator while the lack of gauge invariance leads to the Hamiltonian of the system becomes very complicated. The existence of the fermionic modes propagating faster than "the speed of light" in the minimal coupling to the external electromagnet field is expected to be experimentally examined.   The topological superconductivity of spin-3/2 RS spinors were not studied.  We also anticipate the realization of the RS field in cold atom systems.

\vspace{2mm}

The authors thank Gang Chen, Xi Luo, Yong-Shi Wu and Shiliang Zhu for helpful discussions.  This work is supported by the 973 program of MOST of China (2012CB821402), NNSF of China (11174298, 11474061).


\begin{thebibliography}{99}
\bibitem{note1}  Projecting to $J_{eff}=3/2$ means $|d_{xy},\uparrow\rangle+|d_{yz},\downarrow\rangle+i|d_{zx},\downarrow\rangle=0$ and $-|d_{xy},\downarrow\rangle+|d_{yz},\uparrow\rangle-i|d_{zx},\uparrow\rangle=0$. If we identify $c_{1\sigma}=d_{yz,\sigma}$, $c_{2\sigma}=-d_{zx,\sigma}$ and $c_{3\sigma}=d_{xy,\sigma}$, then the above constraint can be written as $\sigma^ic_i=0$.
\end{thebibliography}
\end{document}